\documentclass[11pt]{amsart}
\usepackage{hyperref}

\begin{document}

\title[Electromagnetic Interaction in Unified Field Theory]
{The Origin of the Electromagnetic Interaction in
Einstein's Unified Field Theory with Sources}
\author{S. Antoci}
\address{Dipartimento di Fisica ``A. Volta'', via Bassi 6,
27100 Pavia, Italy}%

\thanks{General Relativity and Gravitation {\bf 23}, 47 (1991).}
\subjclass{}
\keywords{}

\begin{abstract}
Einstein's unified field theory is extended by the addition of
matter terms in the form of a symmetric energy tensor and of two
conserved currents. From the field equations and from the
conservation identities emerges the picture of a
gravoelectrodynamics in a dynamically polarizable Riemannian
continuum. Through an approximate calculation exploiting this
dynamical polarizability it is argued that ordinary
electromagnetism may be contained in the theory.
\end{abstract}
\maketitle

\section{Introduction}
Recently, it has been shown that, if sources are appended in a
certain way to the field equations of Einstein's unified field
theory \cite{Einstein}, the contracted Bianchi identities and the
field equations appear endowed with definite physical meaning. The
theory looks like a gravoelectrodynamics in a polarizable
Riemannian continuum \cite{Antoci1990}, in which the relationship
between the electromagnetic inductions and fields is dictated by
the field equations in a way that deserves a thorough scrutiny,
for the wealth of the implied possibilities is far richer than in
the so-called Einstein-Maxwell theory. As a partial contribution
to the understanding of these new opportunities, we show here a
way by which the particular occurrence that we consider ordinary
electrodynamic interaction stems from the theory.\par
\section{Einstein's gravoelectrodynamics with sources}
On a four-dimensional, real manifold, let $\mathbf{g}^{ik}$ be a
contravariant tensor density with an even part $\mathbf{g}^{(ik)}$
and an alternating one $\mathbf{g}^{[ik]}$:
\begin{equation}\label{1}
\mathbf{g}^{ik}=\mathbf{g}^{(ik)}+\mathbf{g}^{[ik]}.
\end{equation}
We also endow the manifold with a general affine connection
\begin{equation}\label{2}
W^i_{kl}=W^i_{(kl)}+W^i_{[kl]}.
\end{equation}
Out of this affinity we form the Riemann curvature tensor
\begin{equation}\label{3}
R^i_{~klm}(W)=W^i_{kl,m}-W^i_{km,l}
-W^i_{al}W^a_{km}+W^i_{am}W^a_{kl}
\end{equation}
for which two distinct contractions, $R_{ik}(W)=R^p_{~ikp}(W)$ and
$A_{ik}(W)=R^p_{~pik}(W)$ exist \cite{Schroedinger}. But even the
transposed affinity $\tilde{W}^i_{kl}=W^i_{lk}$ shall be taken into
account: from it, the Riemann curvature tensor
$R^i_{~klm}(\tilde{W})$ and its two contractions
$R_{ik}(\tilde{W})$ and $A_{ik}(\tilde{W})$ can be formed as well.
We aim to follow the pattern of general relativity, in which the
Lagrangian density $\mathbf{g}^{ik}R_{ik}$ is considered, but now
any linear combination $\bar{R}_{ik}$ of the four above-mentioned
contractions can be envisaged. A good choice \cite{Borchsenius},
for reasons that will become apparent later, is
\begin{equation}\label{4}
\bar{R}_{ik}(W)=R_{ik}(W)+\frac{1}{2}A_{ik}(\tilde{W}).
\end{equation}
Since we do not believe that we will attain a complete theory,
i.e. one in which the energy tensor and the currents that describe
matter are determined by the field equations, we endow the theory
with sources in the form of a nonsymmetric tensor $P_{ik}$ and of
a current density $\mathbf{j}^i$, and we couple them to
$\mathbf{g}^{ik}$ and to the vector $W_i=W^l_{[il]}$ respectively.
The Lagrangian density
\begin{equation}\label{5}
\mathbf{L}=\mathbf{g}^{ik}\bar{R}_{ik}(W)
-8\pi\mathbf{g}^{ik}P_{ik}
+\frac{8\pi}{3}W_i\mathbf{j}^i
\end{equation}
is thus arrived at.  By performing independent variations of the
action $\int\mathbf{L}d\Omega$ with respect to $W^p_{qr}$ and to
$\mathbf{g}^{ik}$ with suitable boundary conditions we obtain the
field equations
\begin{eqnarray}\label{6}
-\mathbf{g}^{qr}_{,p}+\delta^r_p\mathbf{g}^{(sq)}_{,s}
-\mathbf{g}^{sr}W^q_{sp}-\mathbf{g}^{qs}W^r_{ps}\\\nonumber
+\delta^r_p\mathbf{g}^{st}W^q_{st}
+\mathbf{g}^{qr}W^t_{pt}
=\frac{4\pi}{3}(\mathbf{j}^r\delta^q_p-\mathbf{j}^q\delta^r_p)
\end{eqnarray}
and
\begin{equation}\label{7}
\bar{R}_{ik}(W)=8\pi P_{ik}.
\end{equation}
By contracting eq. (\ref{6}) with respect to $q$ and $p$ we get
\begin{equation}\label{8}
\mathbf{g}^{[is]}_{,s}={4\pi}\mathbf{j}^i,
\end{equation}
a desirable outcome. We get a problem too, since the very
existence of the latter equation means that we cannot determine
the affinity $W^i_{kl}$ uniquely in terms of $\mathbf{g}^{ik}$:
eq. (\ref{6}) is in fact invariant under the projective
transformation ${W'}^i_{kl}=W^i_{kl}+\delta^i_k\lambda_l$, with
$\lambda_l$ arbitrary vector field. And moreover eq. (\ref{7})
is invariant under the transformation
\begin{equation}\label{9}
{W'}^i_{kl}=W^i_{kl}+\delta^i_k\mu_{,l}
\end{equation}
where $\mu$ is an arbitrary scalar. Equation (\ref{8}) and the
invariance under (\ref{9}) are hints for a possible
electromagnetic content of the theory. We can write
\begin{equation}\label{10}
W^i_{kl}=\Gamma^i_{kl}-\frac{2}{3}\delta^i_kW_l
\end{equation}
where $\Gamma^i_{kl}$ is another affine connection, by definition
constrained to yield $\Gamma^l_{[il]=0}$. Then eq. (\ref{6})
becomes
\begin{equation}\label{11}
\mathbf{g}^{qr}_{,p}+\mathbf{g}^{sr}\Gamma^q_{sp}+\mathbf{g}^{qs}\Gamma^r_{ps}
-\mathbf{g}^{qr}\Gamma^t_{(pt)}
=\frac{4\pi}{3}(\mathbf{j}^q\delta^r_p-\mathbf{j}^r\delta^q_p)
\end{equation}
and allows us to determine $\Gamma^i_{kl}$ uniquely, under very
general conditions, in terms of $\mathbf{g}^{ik}$. When eq.
(\ref{10}) is substituted in eq. (\ref{7}), the latter comes to
read
\begin{eqnarray}\label{12}
\bar{R}_{(ik)}(\Gamma)=8\pi P_{(ik)}\\\label{13}
\bar{R}_{[ik]}(\Gamma)
=8\pi P_{[ik]}-\frac{1}{3}(W_{i,k}-W_{k,i})
\end{eqnarray}
after splitting the even and the alternating parts. Wherever the
source term is nonvanishing, a field equation loses its meaning,
and becomes a definition of some property of matter in terms of
geometrical entities; it is quite obvious that such a definition
must be unique. This occurs with eqs. (\ref{8}), (\ref{11}) and
(\ref{12}), but it does not happen for eq. (\ref{13}). This
equation only prescribes that $\bar{R}_{[ik]}(\Gamma)-8\pi P_{[ik]}$
is the curl of the arbitrary vector $W_i/3$; it is equivalent to
the four equations
\begin{equation}\label{14}
\bar{R}_{[[ik],l]}(\Gamma)=8\pi P_{[[ik],l]}
\end{equation}
and cannot specify $P_{[ik]}$ uniquely. We therefore scrap the
redundant tensor $P_{[ik]}$, as we scrapped the redundant affinity
$W^i_{kl}$, and assume henceforth that matter is defined by the
symmetric tensor $P_{(ik)}$, by the current density $\mathbf{j}^i$
and by the current
\begin{equation}\label{15}
K_{ikl}=\frac{1}{8\pi}\bar{R}_{[[ik],l]},
\end{equation}
both $\mathbf{j}^i$ and $K_{ikl}$ being conserved quantities by
definition. The analogy with general relativity, to which the
present theory formally reduces when $\mathbf{g}^{[ik]}=0$,
suggests rewriting  eq. (\ref{12}) as
\begin{equation}\label{16}
\bar{R}_{(ik)}(\Gamma)=8\pi(T_{ik}
-\frac{1}{2}s_{ik}s^{pq}T_{pq})
\end{equation}
where $s_{ik}=s_{ki}$ is the still unchosen metric tensor of the
theory, $s^{il}s_{kl}=\delta^i_k$, and the symmetric tensor
$T_{ik}$ will act as energy tensor.\par
    Equations (\ref{11}), (\ref{16}), (\ref{8}) and (\ref{15})
reduce to the equations of Einstein's unified field theory when
sources are absent, since then
$\bar{R}_{ik}(\Gamma)$=${R}_{ik}(\Gamma)$; moreover they enjoy
the property of transposition invariance even when sources are
present. If $\mathbf{g}^{ik}$, $\Gamma^i_{kl}$,
$\bar{R}_{ik}(\Gamma)$ represent a solution with the sources
$T_{ik}$, $\mathbf{j}^i$ and $K_{ikl}$, the transposed quantities
$\tilde{\mathbf{g}}^{ik}=\mathbf{g}^{ki}$,
$\tilde{\Gamma}^i_{kl}=\Gamma^i_{lk}$ and
$\bar{R}_{ik}(\tilde{\Gamma})$= $\bar{R}_{ki}(\Gamma)$ represent
another solution, endowed with the sources
$\tilde{T}^{ik}=T_{ik}, \tilde{\mathbf{j}}^i=-\mathbf{j}^i$
and $\tilde{K}_{ikl}=-K_{ikl}$. Such a desirable property is a
consequence of the choice made for $\bar{R}_{ik}$. These equations
suggest interpreting Einstein's unified field theory with sources
as a gravoelectrodynamics in a polarizable continuum, allowing for
both electric and magnetic currents. The study of the conservation
identities confirms the idea and provides at the same time the
identification of the metric tensor $s_{ik}$. Let us consider the
invariant integral
\begin{equation}\label{17}
I=\int\left[\mathbf{g}^{ik}\bar{R}_{ik}(W)
+\frac{8\pi}{3}W_i\mathbf{j}^i\right]d\Omega.
\end{equation}
From it, when eq. (\ref{6}) is assumed to hold, by means of an
infinitesimal coordinate transformation we get the four identities
\begin{eqnarray}\label{18}
-(\mathbf{g}^{is}\bar{R}_{ik}(W)
+\mathbf{g}^{si}\bar{R}_{ki}(W))_{,s}
+\mathbf{g}^{pq}\bar{R}_{pq,k}(W)\\\nonumber
+\frac{8\pi}{3}\mathbf{j}^i(W_{i,k}-W_{k,i})=0.
\end{eqnarray}
This equation can be rewritten as
\begin{eqnarray}\label{19}
-2(\mathbf{g}^{(is)}\bar{R}_{(ik)}(\Gamma))_{,s}
+\mathbf{g}^{(pq)}\bar{R}_{(pq),k}(\Gamma)\\\nonumber
=2\mathbf{g}^{[is]}_{,s}\bar{R}_{[ik]}(\Gamma)
+\mathbf{g}^{[is]}\bar{R}_{[[ik]_{,s}]}(\Gamma)
\end{eqnarray}
where the redundant variable $W^i_{kl}$ no longer appears. Let us
remember eq. (\ref{16}) and assume that the metric tensor is
defined by the equation \cite{Hely}
\begin{equation}\label{20}
\sqrt{-s}s^{ik}=\mathbf{g}^{(ik)}
\end{equation}
where $s=\det{(s_{ik})}$; we shall use henceforth $s^{ik}$ and
$s_{ik}$ to raise and lower indices, $\sqrt{-s}$ to produce
tensor densities out of tensors. We define then
\begin{equation}\label{21}
\mathbf{T}^{ik}=\sqrt{-s}s^{ip}s^{kq}T_{pq}
\end{equation}
and the weak identities (\ref{19}), when all the field equations
hold, will take the form
\begin{equation}\label{22}
\mathbf{T}^{ls}_{;s}=\frac{1}{2}s^{lk}
(\mathbf{j}^i\bar{R}_{[ki]}(\Gamma)
+K_{iks}\mathbf{g}^{[si]})
\end{equation}
where the semicolon indicates the covariant derivative with
respect to the Christoffel affinity
\begin{equation}\label{23}
\left\{^{~i}_{k~l}\right\}
=\frac{1}{2}s^{im}(s_{mk,l}+s_{ml,k}-s_{kl,m})
\end{equation}
built with $s_{ik}$. Our earlier impression is confirmed by eq.
(\ref{22}): the theory, built in terms of a non-Riemannian
geometry, entails a gravoelectrodynamics in a dynamically
polarized Riemannian spacetime, for which $s_{ik}$ is the metric.
The relationship between electromagnetic inductions and fields is
governed by the field equations in a quite novel and subtle way,
with respect to the one prevailing in the so-called
Einstein-Maxwell theory. Two versions of Einstein's
gravoelectrodynamics are possible, according to whether
$\mathbf{g}^{ik}$ is chosen to be a real nonsymmetric or a complex
Hermitian tensor density.
\section{The origin of the electromagnetic interaction}
In the present theory, finding exact solutions to the field
equations not restricted by symmetry or by some other limitation
is by no means easy; even the straightforward task of writing the
sources explicitly in terms of a general $\mathbf{g}^{ik}$ leads
to unsurveyable expressions. We therefore bow to the need to
perform approximate calculations, and assume that, while
$\mathbf{s}^{ik}$ is an arbitrary tensor density,
$\mathbf{g}^{[ik]}$, when compared to $\mathbf{s}^{ik}$, is a
small, first order quantity, that we call henceforth
$\mathbf{a}^{ik}$. As previously noted, we raise and lower indices
with $s^{ik}$ and $s_{ik}$, build tensor densities with
$\sqrt{-s}$, etc. Thus we have
\begin{eqnarray}\label{24}
a_i^{~k}=s_{il}\mathbf{a}^{lk}/\sqrt{-s}=-a^k_{~i}\\\nonumber
a_{ik}=a_i^{~l}s_{kl}=-a_{ki}\\\nonumber
j_i=(1/4\pi)a_{i~;s}^{~s}.
\end{eqnarray}
The structure of the field equations and of the conservation
identities is such that a consistent approximation scheme is
attained if we calculate the alternating quantities in first
order, the even quantities up to second order in $a_{ik}$. Due to
the invariance under transposition, $a_{ik}$ and its derivatives
will appear in the even quantities only through second order
combinations. With this proviso, we solve eq. (\ref{11}) for
$\Gamma^i_{kl}$ up to second order; the approximate solution reads
\begin{equation}\label{25}
\Gamma^i_{kl}=\left\{^{~i}_{k~l}\right\}
+\Theta^i_{kl}+\Sigma^i_{kl}
\end{equation}
where the Christoffel symbol of eq. (\ref{23}) is the zeroth order
contribution, while $\Theta^i_{kl}=-\Theta^i_{lk}$ is the first
order part, and $\Sigma^i_{kl}=\Sigma^i_{lk}$ is the second order
correction. We get
\begin{equation}\label{26}
\Theta^i_{kl}=\frac{1}{2}s^{im}(a_{km;l}+a_{ml;k}-a_{lk;m})
+\frac{4\pi}{3}(\delta^i_k j_l-\delta^i_l j_k)
\end{equation}
and
\begin{eqnarray}\label{27}
\Sigma^i_{kl}=\frac{1}{2}\left\{a^{si}(a_{ks;l}+a_{ls;k})
+a_k^{~s}a^i_{~l;s}+a_l^{~s}a^i_{~k;s}\right\}\\\nonumber
+\frac{1}{4}a^{ms}(\delta^i_ka_{ms;l}
+\delta^i_la_{ms;k}-s^{ip}s_{kl}a_{ms;p})\\\nonumber
+\frac{2\pi}{3}j_s(\delta^i_la_k^{~s}+\delta^i_ka_l^{~s}
-3s_{kl}a^{is}).
\end{eqnarray}
Let us call $S^i_{~klm}$ the Riemann tensor built with
$\left\{^{~i}_{k~l}\right\}$, $S_{ik}$ the corresponding Ricci
tensor. When $\bar{R}_{(ik)}(\Gamma)$ is calculated up to second
order we find
\begin{eqnarray}\label{28}
\bar{R}_{(ik)}(\Gamma)=S_{ik}+\Sigma^a_{ik;a}
-\frac{1}{2}\left(\Sigma^a_{ia;k}+\Sigma^a_{ka;i}\right)
-\Theta^a_{ib}\Theta^b_{ak}\\\label{29}
\bar{R}_{[ik]}(\Gamma)=\Theta^a_{ik;a}
\end{eqnarray}
for the even and the alternating parts respectively. We wish to
understand how the field equations rule the relationship between
$\mathbf{a}^{ik}$ and $\bar{R}_{[ik]}$, and what sort of
interactions arise from the right-hand side of eq. (\ref{22}) when
$\mathbf{j}^i$ and $K_{ikl}$ are everywhere vanishing, except in
preassigned world tubes. In particular, we wish to appreciate in
what manner, if any, the usual electrodynamic interaction is an
outcome of the theory. We need not require that the energy tensor
$T_{ik}$ be vanishing outside the world tubes; we know in advance
from eq. (\ref{22}) that, wherever the two currents annihilate,
$T_{ik}$ displays a pure gravito-inertial behaviour in the
Riemannian spacetime for which $s_{ik}$ is the metric. It will
suffice to solve eqs. (\ref{8}) and (\ref{15}) with the sources
$\mathbf{j}^i$ and $K_{ikl}$ localized in the above sense. When
written in terms of $s_{ik}$ and $a_{ik}$,
$\bar{R}_{[ik]}(\Gamma)$ turns out to be
\begin{eqnarray}\label{30}
\bar{R}_{[ik]}(\Gamma)=\frac{2\pi}{3}(j_{i,k}-j_{k,i})
+\frac{1}{2}a_i^{~n}S_{nk}-\frac{1}{2}a_k^{~n}S_{ni}
\\\nonumber
-a^{pq}S_{pikq}+\frac{1}{2}s^{pq}a_{ik;p;q}.
\end{eqnarray}
Given the form of $\bar{R}_{[ik]}(\Gamma)$, a general solution of
our problem seems beyond reach, but a particular solution, of
immediate physical interest, is at hand. Let us assume that
$a_{ik}$ is a pure curl:
\begin{equation}\label{31}
a_{ik}=\phi_{k,i}-\phi_{i,k}
\end{equation}
everywhere. Then $a_{ik}$ has to obey both sets of Maxwell's
equations in the Riemannian spacetime described by $s_{ik}$; the
vector potential $\phi_i$ shall be so chosen as to ensure the
fulfillment of the equation $a^{~s}_{i~;s}=0$ everywhere except
for preassigned world tubes. Due to eq. (\ref{31})
$\bar{R}_{[ik]}(\Gamma)$ then reads
\begin{equation}\label{32}
\bar{R}_{[ik]}(\Gamma)=\frac{8\pi}{3}(j_{i,k}-j_{k,i})
+a_i^{~n}S_{nk}-a_k^{~n}S_{ni}
+a^{pq}S_{pqik}.
\end{equation}
Remember now the definition of the conformal curvature tensor
\cite{Goldberg}
\begin{eqnarray}\label{33}
C_{ijkl}=S_{ijkl}-\frac{1}{2}\left(S_{jk}s_{il}+s_{jk}S_{il}
-S_{jl}s_{ik}-s_{jl}S_{ik}\right)\\\nonumber
+\frac{S}{6}\left(s_{jk}s_{il}-s_{jl}s_{ik}\right)
\end{eqnarray}
where $S=s^{ik}S_{ik}$ is the scalar curvature of $S^i_{~klm}$;
eq. (\ref{32}) can be rewritten as
\begin{equation}\label{34}
\bar{R}_{[ik]}(\Gamma)=\frac{8\pi}{3}(j_{i,k}-j_{k,i})
+a^{pq}C_{pqik}+\frac{S}{3}a_{ik}.
\end{equation}
Assume that $C_{iklm}=0$ everywhere, i.e. that the Riemannian
spacetime described by $s_{ik}$ is conformally flat, and that $S$
vanishes outside the world tubes. Since $\bar{R}_{[ik]}(\Gamma)$
is a first order quantity, these requirements need to be met in
zeroth order only. Then, to the required order,
$\bar{R}_{[ik]}(\Gamma)$ vanishes outside the world tubes, while
inside it reads
\begin{equation}\label{35}
\bar{R}_{[ik]}(\Gamma)=\frac{8\pi}{3}(j_{i,k}-j_{k,i})
+\frac{S}{3}a_{ik}.
\end{equation}
We imagine now that the world tubes are so chosen that the
intersection of any one of them with an arbitrary spacelike
hypersurface can be individually surrounded by a closed and
otherwise arbitrary two-surface entirely lying where
$\bar{R}_{[ik]}(\Gamma)=0$, and we consider what sort of
interactions are dictated, under these conditions, by the
right-hand side of eq. (\ref{22}). We see that, although $K_{ikl}$
is in general nonvanishing inside a world tube, Gauss theorem
proves that this current cannot give rise to net charges, for
$\bar{R}_{[ik]}(\Gamma)$ vanishes on the two-surface whose
existence was supposed above; only multipole interactions can be
expected from the second term at the right-hand side of eq.
(\ref{22}). From the first term we get two contributions: one is a
current-current self-interaction that vanishes when the curl of
$j_i$ is zero; the other one, which we call $\mathbf{f}^l$, reads
\begin{equation}\label{36}
\mathbf{f}^l=\frac{S}{6}\mathbf{j}_ia^{li}.
\end{equation}
Since $a_{ik}$ obeys both sets of Maxwell's equations, $\mathbf{f}^l$
is a Lorentz force density acting on the current $Sj_i$. If we
assume that the gradient of S is orthogonal to $j^i$, so that the
scalar curvature has constant value along a streamline of $j^i$,
the correspondence with the ordinary electrodynamic interaction
becomes complete, since the current  $Sj_i$ is then a conserved
quantity that can build conserved net charges, like $j_i$ does.
The sign of the interaction can be adjusted to meet the
experimental evidence both in the real nonsymmetric and in the
complex Hermitian versions of the theory by properly choosing the
sign of $S$.
\section{Concluding remarks}
We have provided, through an approximate calculation, evidence of
how the usual electrodynamic interaction can stem from Einstein's
gravoelectrodynamics. We feel confident, after this result, in
identifying $\mathbf{j}^i$ with the electric current density and
$K_{ikl}$ with the magnetic current, $\mathbf{g}^{[ik]}$ with the
electric induction and the magnetic field,
$\bar{R}_{[ik]}(\Gamma)$ with the electric field and the magnetic
induction. Due to the polarizability properties of the continuum
it is not required that, in order to produce the electrodynamic
interaction, a propagating $\bar{R}_{[ik]}(\Gamma)$ should
correspond to each propagating $\mathbf{g}^{[ik]}$. A charged
particle with the appropriate geometrical structure can influence
the polarizability and induce the interaction with its mere
presence.\par
    One should not think, however, that this is the only way the
electrodynamic interaction can arise from the theory: as can
be readily appreciated from eq. (\ref{30}), if $a_{[ik,l]}\neq 0$
and $S^i_{~klm}=0$ everywhere, the field equations ensure, where
currents are absent, that $\bar{R}_{[ik]}(\Gamma)$ must fulfill
the two equations
\begin{eqnarray}\label{37}
(\sqrt{-s}s^{il}s^{km}
\bar{R}_{[lm]}(\Gamma))_{,k}=0\\\label{38}
\bar{R}_{[[ik],l]}(\Gamma)=0.
\end{eqnarray}
Examples of this occurrence with a nonvanishing, propagating
$\bar{R}_{[ik]}(\Gamma)$ have already been provided
\cite{Treder,Antoci1987}. Nor should
one think that the scope of Einstein's gravoelectrodynamics is
limited to accounting for Maxwell's electromagnetism and
gravitation: just in the above-mentioned examples
$\mathbf{g}^{[ik]}$ behaves in such a way that the last term of
eq. (\ref{22}) shows that magnetic charges built by $K_{ikl}$
interact with forces not depending on distance. The interaction of
unlike magnetic charges turns out to be attractive, hence
confining, in the complex Hermitian version of the theory.

\bibliographystyle{amsplain}

\end{document}